\documentclass{article}[12pt]
\usepackage{amssymb}
\def\thefootnote{\fnsymbol{footnote}}
\begin{document}
\begin{titlepage}
\today          \hfill
\begin{center}
%\hfill    RI- \\
\hfill hep-th/yymmxxx  \\

\vskip .5in
\renewcommand{\thefootnote}{\fnsymbol{footnote}}
{\Large \bf  g-function in perturbation theory}
\footnote{This work was supported in part by BSF-American-Israel Bi-National Science Foundation,
the Israel Academy of Sciences and Humanities-Centers of Excellence Program, the German-Israel
Bi-National Science Foundation.
 }
\vskip .50in

\vskip .5in
{\large Anatoly Konechny}\footnote{email address: tolya@phys.huji.ac.il}

\vskip 0.5cm
{\em Racah Institute of Physics\\
The Hebrew University\\
Jerusalem  91904, Israel}
\end{center}

\vskip .5in

\begin{abstract} \large
%insert abstract here
We present some explicit computations checking a particular form of gradient formula for a boundary beta
function in two-dimensional quantum field theory on a disk. The form of the potential function and metric
that we consider were  introduced in \cite{Witt2}, \cite{Shat2} in the context of background independent
open string field theory. We check the gradient formula to the third order in perturbation theory around a
fixed point.  Special consideration is given to
  situations when  resonant terms are present
exhibiting  logarithmic divergences and  universal nonlinearities
in  beta functions. The gradient formula is found to work to the
given order.

\end{abstract}
\end{titlepage}
\large

\newpage
\renewcommand{\thepage}{\arabic{page}}
\setcounter{page}{1}

%%%%%%%%%%%
\large
\section{Introduction}
This paper is devoted to the study of  gradient property of boundary RG flows
in two-dimensional quantum field theory on a surface with boundary.
The gradient property means that there exists  a metric $G_{ij}$ on the space of
boundary conditions and a potential function $g$ such that
\begin{equation}\label{grad_f}
G_{ij}\beta^{j}=-\frac{\partial g}{\partial \lambda^{i}}
\end{equation}
where $\beta^{i}$ is a beta function of  boundary coupling constant $\lambda^{i}$.

In the context of bulk RG flows in 2d theories such a gradient
property was proved by A.~B.~Zamolodchikov in \cite{c-th} (see also \cite{Z_fp},
\cite{Cardy_Ludwig}). He
showed that there exists a potential function  $c$ that is
constant at a fixed point with the value given by the central
charge and that monotonically decreases along the RG flow.
Moreover a concrete construction of such $c$-function and the
metric such that
$$
G_{ij}\beta^{j}=-\frac{\partial c}{\partial \alpha^{i}}
$$
was given in \cite{c-th}. Here $\alpha^{i}$'s are  bulk coupling constants.

For the boundary RG flows a similar statement was conjectured in
\cite{AL1} that goes in the literature under the name "g-theorem".
A  perturbative proof of the conjecture was presented in
\cite{AL2}. The number analogous to the central charge that is
supposed to decrease from UV to IR fixed point is called a
boundary entropy $g$ and at a fixed point it is defined as follows.
Consider a quantum critical system on a cylinder of length $l$ and circumference $r$.
A conformally invariant boundary codition at the ends of the cylinder can be
represented by a boundary state $|B\rangle$ \cite{Cardy}.
In the limit of large $l$ the cylinder partition function has asymptotics
\begin{equation}\label{cylinder_pf}
Z_{BB}(l, r) = \langle B|e^{-lH}|B\rangle \, \sim \, \langle
B|0\rangle\langle 0|B\rangle e^{-E_{0}l}
\end{equation}
where $|0\rangle$ is a vacuum state
for periodic boundary condition on a cylinder and $E_{0}$ is the ground state energy.
 The boundary entropy
is then defined as a number
$$
g=\langle B|0\rangle=Z_{disk}
$$
that equals the value of the disk partition function. More precisely the boundary state is normalized by
equating open (strip) and closed string (cylinder) channel representations for the cylinder partition
function  \cite{AL1}, \cite{least_act}. (The phase of $|B\rangle$ can be chosen so that $\langle
B|0\rangle$ is real and positive and thus $g$ is such as well.)

The perturbative computation presented in \cite{AL2}  in essence
goes as follows. The authors consider a theory on a semi-infinite
strip perturbed by a single boundary primary operator $\phi$ of
dimension $\Delta=1-\epsilon$ with $0<\epsilon << 1$. A Kosterlitz
type  renormalization scheme (see \cite{Kosterlitz},
\cite{Ludwig}, \cite{Cardy_lect}) is chosen in which the
corresponding beta function is
\begin{equation} \label{Kbeta}
\beta(\lambda) = \epsilon\lambda + C\lambda^{2}
\end{equation}
where $C$ is the OPE coefficient of $\phi$ with itself:
$$
\phi(\tau)\phi(0)\sim \frac{C\phi(0)}{|\tau|^{1-\epsilon}} \, .
$$
We note that the quadratic term in this beta function is scheme dependent. For example by a suitable
coupling constant redefinition one can make the beta function linear. The IR  fixed point present in
(\ref{Kbeta}) is then pushed to infinity. Employing the scheme that gives (\ref{Kbeta}) allows one
to
compare the values of the
 partition function at the nearby fixed points. By carefully dropping terms extensive
 in the strip width and performing a proper renormalization
 the authors of \cite{AL2}
 arrive at the change in boundary entropy between the fixed points
\begin{equation} \label{delta_g}
\frac{\delta g}{g}=-\frac{\pi^{2}\epsilon^{3}}{3C^{2}}
\end{equation}
exact to the order $\epsilon^{3}$.

This formula proved to be  useful in the analysis of examples of boundary flows (see e.g. \cite{RRS},
\cite{GRW} and references therein). However its derivation does not provide us with the potential function
and metric of the gradient formula. Also it would be desirable to extend the perturbative analysis to more
general flows whith several   coupling constants running.

Note that   the gradient property itself
is easy to establish to the third order in perturbation theory.
 In a certain renormalization  scheme \cite{Cardy_lect}
the beta functions
are
\begin{equation}
\beta^{i} = \frac{d \lambda^{i}}{d\ln L}=
\epsilon_{i}\lambda_{i} + \sum_{jk}C_{i(jk)}\lambda^{j}\lambda^{k}
\end{equation}
where $\epsilon_{i}=1-\Delta_{i} $, $C_{i(jk)} = \frac{1}{2}(C_{ijk}+C_{ikj})$ and $C_{ijk}$ are the
boundary 3-point structure constants. Here $L$ is the position space renormalization scale. The gradient
property then follows from the fact that $C_{i(jk)}$ is totally symmetric in its indices that in its turn
follows from the cyclic symmetry of $C_{ijk}$. Our interest though is in a canonical  form of the metric
and potential  function that would work to all orders similar to the ones that were constructed by
A.~B.~Zamolodchikov for the bulk beta functions.

A concrete proposal for such a  form came about in \cite{Witt2}, \cite{Shat2} in the framework of
background independent open string field theory that was put forward in \cite{Witt1}. The tentative
potential function   has the form
\begin{equation} \label{g}
g=Z - \sum_{i}\beta^{i}\frac{\partial Z}{\partial \lambda^{i}}
\end{equation}
where
$$
Z=\int [d\phi]e^{-S}
$$
is the renormalized disk partition function, while the metric is
\begin{equation} \label{metric}
G_{ij}=\int\limits_{0}^{2\pi}\frac{rd\theta_{1}}{2\pi}\int\limits_{0}^{2\pi}\frac{rd\theta_{2}}{2\pi}
\langle \phi_{i}(re^{i\theta_{1}})\phi_{j}(re^{i\theta_{2}})\rangle 2\sin^{2}
\left(\frac{\theta_{1}-\theta_{2}}{2}\right) \, .
\end{equation}
Here $\langle ... \rangle$ stands for a nonnormalized correlator that does not include the division by
$Z$, $r$ is the radius of the disk.

 Formula (\ref{g}) in its
covariant form as well as its analysis in conformal perturbation
theory first appear in \cite{Shat2}. We find though the
alternative computations done in the present paper more explicit.
Also we make no reference in our analysis to string theory objects
such as conformal ghosts and BRST operator dealing only with field
theoretic structures.
 We hope that this approach may be more beneficiary in clarifying the field theoretic status
of the gradient formula (\ref{grad_f}), (\ref{g}), (\ref{metric}).

The approach of \cite{Witt1}, \cite{Witt2}, \cite{Shat1}, \cite{Shat2} was
successfully applied not long ago
to the analysis of tachyon condensation \cite{GS}, \cite{KMM}. A particular exactly
calculable model \cite{Witt2}
with quadratic perturbation and linear beta functions
was considered along the way and it was shown in
\cite{KMM} that (\ref{g}) monotonically decreases along the RG flow for that model.
The model considered in \cite{GS}, \cite{KMM} is a sigma model with noncompact
target space. A noncompact case  in general
 may require a  special care.
For instance it is  known that the $c$-theorem can fail in the noncompact situation \cite{Polch} (see
\cite{panic} for a recent discussion of such situation). The same concerns application of formulae
(\ref{g}), (\ref{metric}) in the noncompact case. Thus in \cite{KMM} one of the coordinates was regarded
as a space-time volume regulator and was treated in a special way.

In the boundary case one concrete  problem in noncompact situation (irrational CFT) arises regarding the
value of boundary entropy at a fixed point. For example in the case when we consider a sigma model whose
target space is noncompact and translation invariant one may be tempted to define $g$ by dividing
$Z_{disk}$ over the infinite space-time volume $V$. It that case however the value of $Z_{disk}/V$ may
depend on the value of an exactly marginal boundary coupling as happens for instance in the case of
constant $U(1)$ field strength model (\cite{FT}, \cite{Callan_etal}). And thus in the field theoretic
sense this value cannot be a gradient function\footnote{I am grateful to Daniel Friedan for the discussion
on this point} (although it still may define a string space-time effective  action).

The paper is organized as follows. In section 2 we perform a third order check of the gradient formula
done in the absence of logarithmic divergences. The next three sections deal with cases when low order
resonant terms are present each exhibiting a logarithmic divergence and  universal nonlinearities in beta
functions: second order resonance with nonvanishing linear terms in the beta functions (section 3),
quadratic and qubic resonances in the identity coupling beta function (section 4), marginal but not
exactly marginal couplings (section 5). In section 6 we conclude by pointing out some open questions. The
appendix contains an explicit expression for the first perturbative correction to the local two point
function. Although we do not use it in the analysis of the gradient formula this expression complements
naturally the  perturbative computations done in the paper and may be found of use in the future.

\section{Third order computation in the absence of logarithmic divergences}
We consider a two-dimensional quantum field theory  on a disk $|z|^{2}\le r^{2}$
whose (Euclidean) action functional has the form
\begin{equation} \label{S}
S=S_{0} - \sum_{i} \int_{0}^{2\pi} \frac{rd\theta}{2\pi}
\lambda^{i}\phi_{i}(re^{i\theta}) -\alpha r
\end{equation}
where $S_{0}$ defines a conformal field theory with conformal boundary conditions
and $\phi_{i}$ are its boundary primary fields with weights $\Delta_{i}$.
For simplicity we assume that $\Delta_{i}\ne \Delta_{j}$ if $i\ne j$.
The modifications needed for the more general case are straightforward.
 $\alpha$ is
the coupling constant of identity operator.

Throughout this section we will assume that integrals arising in perturbation series in $\lambda^{i}$'s
are all power divergent up to (and including) the third order.  We will employ a minimal subtraction
scheme. It is easy to show that under these assumptions  all beta functions remain linear to the given
order. In particular the beta function of the identity operator is $\beta_{ \bf 1}=\alpha$. We will treat
the coupling constant $\alpha$ nonperturbatively. Let us  show now that its contribution to the gradient
formula decouples from the rest of coupling constants. The renormalized disk partition function can be
written as
\begin{equation}
Z_{disk} = Z=\int [d\phi]e^{-S}=e^{\alpha r}\tilde Z(r, \lambda) \, .
\end{equation}
Plugging this into the gradient formula we obtain
\begin{eqnarray}\label{id_coupl}
G_{{\bf 1}i}\beta^{i} + G_{\bf 11}\alpha= \alpha r^{2} e^{\alpha r
}\tilde Z +
r e^{\alpha r}\beta^{i}\frac{\partial \tilde Z}{\partial \lambda^{i}} \, , \nonumber\\
G_{ij}\beta^{j}+ G_{i{ \bf 1}}\alpha = -e^{\alpha
r}\frac{\partial} {\partial \lambda^{i}}(\tilde Z -
\beta^{j}\frac{\partial \tilde Z}{\partial \lambda^{j}}) + \alpha
r e^{\alpha r}\frac{\partial\tilde Z} {\partial \lambda^{i}} \, .
\end{eqnarray}
It follows from the definition of the metric (\ref{metric}) that
\begin{eqnarray}
G_{\bf 11} = r^{2}e^{\alpha r}\tilde Z \, , \\
G_{{\bf 1}i}=G_{i{\bf 1}} = re^{\alpha r}\frac{\partial \tilde Z}{\partial \lambda^{i}}
\end{eqnarray}
and thus the first equation in (\ref{id_coupl}) holds identically while the second one
(and therefore the whole gradient formula) boils down to
\begin{equation}
\tilde G_{ij}\beta^{j} = -\frac{\partial}
{\partial \lambda^{i}}(\tilde Z -
\beta^{j}\frac{\partial \tilde Z}{\partial \lambda^{j}})
\end{equation}
where we introduced $\tilde G_{ij}$ that is given by formula
(\ref{metric}) with the factor $e^{\alpha r}$ omitted. To simplify
the notation we will omit below the tilde over $\tilde Z$ and
$\tilde G_{ij}$ but it will be assumed that the factors containing
$\alpha$ are everywhere dropped.
%%%%%%%%%%%%%%%%%%

To organize the perturbation theory expansion it is convenient to label various terms
of order $n$ in $\lambda^{i}$'s by an upper index $(n)$. Thus at the second order
we have to prove that
\begin{equation}\label{2o}
 G_{ij}^{(0)}\beta^{j(1)}=-\partial_{i}( Z^{(2)}-\beta^{j(1)}\partial_{j}
 Z^{(2)}) \,.
\end{equation}
 At this order we encounter an integral
\begin{equation}
I_{2}(\nu)\equiv
\int_{0}^{2\pi}\frac{d\theta}{2\pi}\Bigl[\sin^{2}\left(\frac{\theta}{2}\right)\Bigr]^{\nu} =
\frac{\Gamma(\nu + \frac{1}{2})}{\sqrt{\pi}\Gamma(1+\nu)} \, , \quad \nu >-\frac{1}{2}
\end{equation}
that for $\nu \ne -1/2$ is defined via analytic continuation that is equivalent
to dropping the power divergence.
We find then that
\begin{eqnarray}
 Z^{(2)} = \frac{1}{8\sqrt{\pi}}\sum_{i}(\lambda^{i})^{2}(2r)^{2\epsilon_{i}}
\frac{\Gamma(\epsilon_{i}-\frac{1}{2})}{\Gamma(\epsilon_{i})} \, , \label{Z2}\\
 G_{ij}^{(0)} =\frac{(2r)^{2\epsilon_{i}}\delta_{ij}}{2} \label{G0}
\frac{\Gamma(\epsilon_{i}+\frac{1}{2})}
{\sqrt{\pi}\Gamma(1+\epsilon_{i})} \, .
\end{eqnarray}
Plugging these expressions into the both sides of (\ref{2o}) we find that the equality indeed
holds. This computation was also done in \cite{KMM}.
Note that the pole at $\epsilon_{i}=1/2$ that is present in $ Z^{(2)}$ (\ref{Z2})
and that corresponds to a logarithmic running of the identity coupling constant
disappears from the $g$ function upon subtracting $\beta^{j(1)}\partial_{j}
 Z^{(2)}$. We will see the same effect at the next order.

At the third order the identity we are supposed to check reads
\begin{equation} \label{3o}
G_{ij}^{(1)}\beta^{j(1)} =
-\partial_{i}(   Z^{(3)} - \beta^{k(1)}\partial_{k} Z^{(3)} )
\end{equation}
A triple correlator of primaries $\phi_{i}$ is fixed by modular invariance
and we encounter the following integral
\begin{eqnarray} \label{I3}
I_{3}(\nu_{1},\nu_{2}, \nu_{3})\equiv
\frac{1}{(2\pi)^{2}}\int_{0}^{2\pi}d\theta_{1}\int_{0}^{2\pi}d\theta_{2}
\Bigl[\sin^{2}\left(\frac{\theta_{1}}{2}\right)\Bigr]^{\nu_{1}}
\Bigl[\sin^{2}\left(\frac{\theta_{2}}{2}\right)\Bigr]^{\nu_{2}}
\Bigl[\sin^{2}\left(\frac{\theta_{1}+\theta_{2}}{2}\right)\Bigr]^{\nu_{3}} =\nonumber \\
 \frac{\Gamma(1+\nu_{1}+\nu_2 + \nu_3)}{\pi^{3/2}}
\frac{\Gamma(\nu_{1}+\frac{1}{2})\Gamma(\nu_{2}+\frac{1}{2})\Gamma(\nu_{3}+\frac{1}{2})}
{\Gamma(\nu_1+\nu_2 + 1)\Gamma(\nu_2+\nu_3 + 1)\Gamma(\nu_1+\nu_3 + 1)}
\end{eqnarray}
that for values $\nu_{i}<-\frac{1}{2}$ is defined via analytic continuation.
A computation leading to (\ref{I3}) can be found e.g. in the Appendix A of \cite{Frolov}.
The integral can be first mapped on the half plane where
we obtain a rational integrand, which in its turn can be integrated by standard
Feynman parameters technique.

For $Z^{(3)}$ we have an expression
\begin{equation}
Z^{(3)}=\frac{1}{3!}
\sum_{ijk}\lambda^{i}\lambda^{j}\lambda^{k}(2r)^{\epsilon_{i}+\epsilon_{j}+
\epsilon_{k}}C_{ijk}K_{ijk}
\end{equation}
where
\begin{eqnarray}\label{K}
K_{ijk} =
\frac{1}{32\pi^{2}}\int_{0}^{2\pi}d\theta_{1}\int_{0}^{2\pi}d\theta_{2}
\Bigl[\sin^{2}\left(\frac{\theta_{1}}{2}\right)\Bigr]^{\frac{1}{2}(\Delta_{i}-\Delta_{j}-
\Delta_{k})} \nonumber \\
\Bigl[\sin^{2}\left(\frac{\theta_{2}}{2}\right)\Bigr]^{\frac{1}{2}(\Delta_{j} - \Delta_{i} - \Delta_{k})}
\Bigl[\sin^{2}\left(\frac{\theta_{1}+\theta_{2}}{2}\right)\Bigr]^{\frac{1}{2}( \Delta_{k} - \Delta_{i} -
\Delta_{j})}
\end{eqnarray}
and $C_{ijk}$ are the OPE coefficients. This integral has subdivergences whenever
\begin{equation} \label{sub}
\epsilon_{i} \ge \epsilon_{j} + \epsilon_{k}
\end{equation}
and an overall divergence if
\begin{equation} \label{over}
\epsilon_{i} + \epsilon_{j} + \epsilon_{k} \le 1 \, .
\end{equation}
Assuming that the equalities in (\ref{sub}), (\ref{over}) do not hold
we can evaluate (\ref{K}) using (\ref{I3}) via analytic continuation
\begin{equation} \label{Z_3}
K_{ijk} =\frac{\Gamma(\frac{1}{2}(-1+\epsilon_{i} + \epsilon_j +
\epsilon_k))} {(4\pi)^{3/2}}\times
\frac{\Gamma(\frac{1}{2}(\epsilon_j+\epsilon_k-\epsilon_i))
\Gamma(\frac{1}{2}(\epsilon_i+\epsilon_k-\epsilon_j))
\Gamma(\frac{1}{2}(\epsilon_i+\epsilon_j-\epsilon_k))}
{\Gamma(\epsilon_i)\Gamma(\epsilon_j)\Gamma(\epsilon_k)} \, .
\end{equation}
The poles in this expression have a natural interpretation in terms of resonances (to be discussed in more
detail in the following sections).
 Thus the poles
$$
\epsilon_{i}=\epsilon_j + \epsilon_k
$$
correspond to a resonance term proportional to $\lambda^{j}\lambda^{k}$ in $\beta^{i}$ while the poles
\begin{equation} \label{1res}
\epsilon_i + \epsilon_j + \epsilon_k = 1
\end{equation}
correspond to  resonance terms proportional to $\lambda^{i}\lambda^{j}\lambda^{k}$  in the beta function
of  identity operator.

The first order correction to the metric has the form
\begin{eqnarray} \label{G1}
&G_{ij}^{(1)} = \sum_{k}\lambda^{k}r^{\epsilon_i + \epsilon_j +
\epsilon_k} \int \frac{d\theta_1}{2\pi}\int
\frac{d\theta_2}{2\pi}\int \frac{d\theta_3}{2\pi} \langle
\phi_i(\theta_{1})\phi_{j}(\theta_{2})\phi_{k}(\theta_{3})\rangle
2\sin^{2}\left(\frac{\theta_{1}-\theta_{2}}{2}\right) =\nonumber \\
&\sum_{k}\lambda^{k}(2r)^{\epsilon_i + \epsilon_j + \epsilon_k}C_{(ijk)}F_{ij}^{k}
\end{eqnarray}
where
\begin{equation} \label{F}
F_{ij}^{k} = \frac{
\Gamma(\frac{1}{2}(1+\epsilon_i + \epsilon_j + \epsilon_k))}{4\pi^{3/2}}
\frac{\Gamma(1+\frac{1}{2}(\epsilon_i + \epsilon_j-\epsilon_k))
\Gamma(\frac{1}{2}(\epsilon_i + \epsilon_k - \epsilon_j))
\Gamma(\frac{1}{2}(\epsilon_j + \epsilon_k - \epsilon_i))}
{\Gamma(1+\epsilon_i)\Gamma(1+\epsilon_j) \Gamma(\epsilon_k)} \, .
\end{equation}
In (\ref{G1})  and everywhere below $C_{(ijk)}$ denotes the symmetrized OPE coefficients.

From the above formulas one easily obtains the relation
$$
F_{ij}^{k}=\frac{1}{2}\Bigl[K_{ijk}\frac{1-\epsilon_i-\epsilon_j-\epsilon_k}
{\epsilon_i \epsilon_j \epsilon_k}\Bigr]\times \epsilon_k (\epsilon_k-\epsilon_i-\epsilon_j) \, .
$$

Using this relation and noting that
$$
\beta^{i(1)} = \epsilon_{i}\lambda^{i}
$$
it is straightforward to derive
\begin{eqnarray}
G_{ij}^{(1)}\beta^{j(1)} &=& -\frac{1}{2}\sum_{jk}(2r)^{\epsilon_i
+ \epsilon_j + \epsilon_k} C_{ijk}K_{ijk}(1-\epsilon_i -
\epsilon_j -\epsilon_k)\lambda^{i}\lambda^{j}\lambda^{k} =
\nonumber \\
&&-\frac{\partial}{\partial \lambda^{i}}
(Z^{(3)} - \beta^{j(1)}\frac{\partial Z^{(3)}}{\partial \lambda^{j}}) \, .
\end{eqnarray}

We see that in the absence of logarithmic divergences the gradient formula
(\ref{grad_f}), (\ref{g}), (\ref{metric}) holds to the third order in perturbation theory.
Note  again that like at the second order the resonance poles (\ref{1res})
corresponding to logarithmic
running of the identity coupling constant were subtracted in the final expression for
$Z^{(3)}$. This looks suggestive of the fact that this effect may happen to all
orders in perturbation theory.

As a final remark in this section let us note that the renormalized
partition function satisfies a simple finite size scaling relation
\begin{equation} \label{f_size}
r\frac{d  Z}{dr} = \beta^{i}\frac{\partial  Z}{\partial \lambda^{i}} \, .
\end{equation}
Thus the $g$ function could be written as
$$
g = Z - r\frac{d  Z}{dr} \, .
$$
We will return to this representation  of $g$-function in the last section.

\section{Resonant terms}
Let us start with a brief reminder of general facts about
perturbation theory resonant terms. Renormalization group
equations in general have a form
\begin{equation} \label{d_eq}
\frac{d \lambda^{i}}{d t} = D^{i}_{j}\lambda^{j} + h^{i}(\lambda)
\end{equation}
where $D$ is the matrix of anomalous dimensions at the fixed point and $h_{i}(\lambda)$ contains all the
nonlinearities. We assume that the coordinates $\lambda^{j}$, $j=1, \dots , n$ in the space of the
theories are chosen so that $D$ is in its Jordan normal form. The  $h^{i}(\lambda)$ can be written as  a
formal power series with a typical term of the form
\begin{equation} \label{mon}
h^{i}_{j_{1}j_{2}...j_{n}}(\lambda^{1})^{j_{1}}\cdot \dots \cdot (\lambda^{n})^{j_{n}}
\end{equation}
where $j_{k}$'s are nonnegative integers.
Different renormalization schemes are related by a formal change of coordinates
$$
{\lambda'}^{i}(\lambda) = \lambda^{i} + \xi^{i}(\lambda) \, .
$$
One may try to choose  $\xi^{i}(\lambda)$ such that the transformed equation (\ref{d_eq})
takes the simplest possible form. In the best case the system (\ref{d_eq}) can be
brought to the linear form
\begin{equation} \label{linear}
\frac{d {\lambda'}^{i}}{d t} = D^{i}_{j}{\lambda'}^{j} \, .
\end{equation}

It is known in the theory of differential equations that obstructions
to linearization of the system (\ref{d_eq}) are the so called resonant monomials.
Let $(\epsilon_{1}, \dots , \epsilon_{n})$ be the set of eigenvalues of the matrix $D$.
A monomial of the form (\ref{mon})  is called resonant if
\begin{equation} \label{resonance}
\sum_{k=1}^{n}\epsilon_{k}j_{k} = \epsilon_{i} \, , \qquad \sum_{k} j_{k} \ge 2 \, .
\end{equation}
In the context of conformal perturbation theory in 2 dimensions (in the bulk or on the boundary)
one can identify coordinates $\lambda^{i}$ with the coupling constants appearing in
(\ref{S}). The matrix $D$ in this case is diagonal.
One can estimate the perturbation expansion divergencies  emerging when points of insertion of several operators
$\phi_{i}$ come together via operator product expansion. One sees then that the resonance condition
(\ref{resonance}) implies a logarithmically divergent counterterm for the $i$-th coupling
constant that is poroportional to the corresponding monomial (\ref{mon}).

If resonant terms are present in the RHS of (\ref{d_eq}) those equations cannot be linearized by any
choice of coordinates. Assuming  that $D={\rm diag}(\epsilon_{1}, \dots , \epsilon_{n})$ one can prove
though (see e.g. \cite{n_forms} chapter 2, theorem 1.5) that there exists a formal change of coordinates
  such that in the new  coordinates the nonlinear parts $h^{i}(\lambda)$ consist of
  resonant monomials.
In the absence of resonant terms there are stronger results available. The theorems of Poincare and Siegel
give sufficient conditions  for the existence of an {\it analytic} change of coordinates that
bring (\ref{d_eq}) to the form (\ref{linear}). We refer the interested reader to the book \cite{n_forms}
for the precise statements of those theorems and more of the mathematical background on normal forms of
differential equations. In conformal perturbation theory the resonant terms were discussed in
\cite{Z_int}, \cite{AlZ}. In the context of background independent string field theory the role of
resonances was emphasized in \cite{Shat2}.

We see thus that although we did a third order computation in the previous section we have not really
tested whether the gradient formula at hand handles universal nonlinearities. To do that we consider in
this and the next two sections situations when low order resonant terms are present.

 Consider now a fixed point perturbed by 3 fields boundary fields $\phi_{i}$, $i=1,2,3$ whose anomalous
dimensions satisfy
 a resonant condition
\begin{equation}
\epsilon_{3} = \epsilon_{1} + \epsilon_{2} \, .
\end{equation}
We will further assume that the only nonvanishing OPE coefficients are $C_{123}$
and the ones with permuted indices. In the point splitting + minimal subtraction scheme
the beta functions  have the form
\begin{eqnarray} \label{betas}
&&\beta^{3} = \epsilon_{3}\lambda^{3} + \frac{1}{\pi}\lambda^{1}\lambda^{2} C_{(123)} + ...\, , \nonumber \\
&&\beta^{1} = \epsilon_{1}\lambda^{1} + ...\, ,\nonumber \\
&&\beta^{2} = \epsilon_{2}\lambda^{2} + ...\, .
\end{eqnarray}
through the third order in the coupling constants. The quadratic term in the first equation
in (\ref{betas}) is universal.

The integral expressions for $K_{123}$ and $F_{13}^{2}=F_{31}^{2}$ (\ref{K}), (\ref{F})
contain a logarithmic
divergence. If $\delta$ is a cutoff of the angular variable $\theta$ the divergences are
proportional to $-2\ln \delta$ that in terms of dimensionfull position space cutoff $a$
can be written as $2\ln(r/a)$. We will employ a  minimal type subtraction scheme in which
this divergence is subtracted with an additional finite term of the form $2\ln(r\mu)$
where $\mu$ is a renormalization mass scale.

%%%%%%%%%%%%%%%
In this scheme we obtain
\begin{equation}
(Z^{(3)})_{ren}= \lambda^{1}\lambda^{2}\lambda^{3}(2r)^{2\epsilon_{1}+2\epsilon_{2}}
C_{(123)}\frac{\Gamma(\epsilon_{1} + \epsilon_{2} -\frac{1}{2})}
{8\pi^{3/2}\Gamma(\epsilon_{1}+\epsilon_{2})} (2\ln(r\mu) + g_{12})
\end{equation}
where
$$
g_{12} \equiv \psi(\epsilon_{1} + \epsilon_{2} - \frac{1}{2}) -\psi(\epsilon_{1}) -
\psi(\epsilon_{2})
$$
where $\psi(x)=\frac{d\ln \Gamma(x)}{dx}$ is the logarithmic derivative of the Euler's
Gamma function. To obtain this formula one should expand the integral
$I_{3}$ (\ref{I3}) in $\nu_{3}$ around $\nu_{3}=-\frac{1}{2}$ keeping
$\nu_{1}=\epsilon_{2} - 1/2$ and $\nu_{2}=\epsilon_{1}-1/2$ fixed.
Then take the limit $\nu_{3}\to -\frac{1}{2}$
 subtracting the pole.

Similarly one obtains
\begin{equation}
(F_{13}^{2})_{ren}=\frac{\Gamma(\epsilon_{1}+\epsilon_{2}
+\frac{1}{2})}{4\pi^{3/2}\Gamma(\epsilon_{1}+\epsilon_{2}+1)}
\left(2\ln(r\mu) + g_{12} + \frac{1}{\epsilon_{1} + \epsilon_{2} -1/2} -\frac{1}{\epsilon_{1}}
\right)
\end{equation}
and the same kind of expression for $(F_{23}^{1})_{ren}$ with $\epsilon_{1}$ interchanged with
$\epsilon_{2}$.

Note that we can use the renormalization scale $\mu$ to introduce dimensionless couplings
$\tilde \lambda^{i}=\mu^{-\epsilon_{i}} \lambda^{i}$. It is easy to check then that the
renormalized partition function $Z_{ren} = 1 + Z^{(2)} + (Z^{(3)})_{ren}+ ...$
up to the third order satisfies the following
 renormalization group equation
 $$
\mu \frac{\partial Z_{ren}(\tilde \lambda, \mu)}{\partial \mu} = \beta^{i}(\tilde \lambda)
\frac{\partial Z_{ren}(\tilde \lambda, \mu)}{\partial \tilde \lambda^{i}}
 $$
with $\beta^{i}$'s given in (\ref{betas}). Also the simple finite size scaling relation
given in (\ref{f_size}) holds.

The coefficient $F_{12}^{3}$ does not require any renormalization (besides the usual subtraction
of power divergences) and is given by
\begin{equation}
F_{12}^{3}=\frac{\Gamma(\epsilon_{1}+\epsilon_{2}+\frac{1}{2})}
{4\epsilon_{1}\epsilon_{2}\pi^{3/2}\Gamma(\epsilon_{1} + \epsilon_{2} )} \, .
\end{equation}

We are fully equipped now to check the equations
\begin{eqnarray} \label{g_eq}
 && G_{1j}^{(1)}\beta^{j(1)}=\left[ F_{12}^{3}\beta^{2(1)}\lambda^{3} +
(F_{13}^{2})_{ren}\beta^{3(1)}\lambda^{2} \right]C_{(123)}(2r)^{2\epsilon_{3}} =
-\frac{\partial g^{(3)}}{\partial \lambda^{1}} \, , \nonumber\\
 &&G_{2j}^{(1)}\beta^{j(1)}=\left[ F_{21}^{3}\beta^{1(1)}\lambda^{3} +
(F_{23}^{1})_{ren}\beta^{3(1)}\lambda^{1} \right]C_{(123)} (2r)^{2\epsilon_{3}} =
-\frac{\partial g^{(3)}}{\partial \lambda^{2}} \, ,  \nonumber\\
   && G_{3j}^{(1)}\beta^{j(1)} + G_{33}^{(0)}\beta^{3(2)} =
  {[}(F_{32}^{1})_{ren}\beta^{2(1)}\lambda^{1} +
(F_{31}^{2})_{ren}\beta^{1(1)}\lambda^{2} {]}C_{(123)}(2r)^{2\epsilon_{3}}  + \nonumber
  \\
 && + \, G_{33}^{(0)}\beta^{3(2)}  =-\frac{\partial g^{(3)}}{\partial
\lambda^{3}}
\end{eqnarray}
where
$$
g^{(3)} = (Z^{(3)})_{ren} - \sum_{i=1}^{3}\beta^{i(1)}\frac{\partial (Z^{(3)})_{ren}}
{\partial \lambda^{i}} - \beta^{3(2)}\frac{\partial Z^{(2)}}
{\partial \lambda^{3}} \, .
$$
A straightforward computation shows that  the equations (\ref{g_eq}) indeed hold.
\section{Resonances in the identity coupling beta function}
So far we considered the cases when the beta function of the
identity operator is exactly linear: $\beta_{\bf 1}=\alpha$. In
this section we will consider two cases when $\beta_{\bf 1}$ is of
a more general form
\begin{equation}
\beta_{\bf 1} = \alpha + h(\lambda^{i}) \, .
\end{equation}
In this case one of the gradient equations
\begin{equation} \label{grad_alpha}
G_{{\bf 1}i}\beta^{i} + G_{\bf 11}\beta_{\bf 1} = -\frac{
\partial g}{\partial \alpha}
\end{equation}
holds identically while the remaining set of equations takes the
form
\begin{equation} \label{grad2}
G_{ij}\beta^{j} = -\frac{\partial}{\partial \lambda^{i}}(Z - \beta^{j}\frac{\partial Z}{\partial
\lambda^{j}}) + rZ\frac{\partial h}{\partial \lambda^{i}}
\end{equation}
Again the factors $e^{\alpha r}$ can be dropped on both sides of
the equation.

 In section 2 we  noted two  resonances of
the identity coupling. One happens at the second order in
perturbation expansion when a coupling of dimension $\Delta =
\frac{1}{2}$ is present while another one takes place at the third
order when we have 3 couplings whose anomalous dimensions satisfy
\begin{equation} \label{res3}
\epsilon_{1} + \epsilon_{2} + \epsilon_{3} = 1 \, .
\end{equation}

In the first case let us restrict our attention only to the identity coupling and a coupling constant
$\lambda$ of the dimension $1/2$ operator. The beta functions then are readily shown to be
\begin{eqnarray*}
&&\beta_{\bf 1} = \alpha + \frac{\lambda^{2}}{2\pi} + \dots \, , \\
&&\beta_{\lambda} = \frac{\lambda}{2} + \dots
\end{eqnarray*}
through the second order in couplings. The expression $(Z^{(2)})_{ren} -
\beta_{\lambda}^{(1)}\partial_{\lambda}(Z^{(2)})_{ren}$ vanishes and the gradient formula (\ref{grad2}) at
the leading order boils down to the equation
$$
G_{\lambda \lambda}^{(0)}\beta_{\lambda}^{(1)} = r\partial_{\lambda} \beta_{\bf 1}^{(2)}
$$
that upon using (\ref{G0}) is readily found to be correct. At the next order in perturbation expansion our
analysis done in section 2 is still valid because the additional term $Z\frac{\partial h}{\partial
\lambda}$ next contributes at the 4th order.

Let us now look at the second case when there are 3 couplings $\lambda^{i}$, $i=1,2,3$ whose anomalous
dimensions satisfy (\ref{res3}). By analyzing the behavior of integral (\ref{K}) in the regions
$\theta_{1,2}\to 0, 2\pi$ we find that it contains a logarithmic divergence $ K_{123}^{*}\ln M$, $M\to
\infty$ where
\begin{eqnarray}
&& K_{123}^{*} = \frac{\Gamma(\nu_{1} + \frac{1}{2})\Gamma(\nu_{2} + \frac{1}{2}) \Gamma(\nu_{3} +
\frac{1}{2})}{8\pi^{3/2}\Gamma(-\nu_{1})\Gamma(-\nu_{2})\Gamma(-\nu_{3})} \, ,\\
&& \nu_{1} = \frac{1}{2}( \epsilon_{2} + \epsilon_{3} - \epsilon_1 -1 )
\end{eqnarray}
and $\nu_{2}$, $\nu_{3}$ are defined by cyclic permutation. The beta functions thus have the form
\begin{eqnarray*}
&& \beta^{i} = \epsilon_{i}\lambda^{i} + \dots \, , \enspace i=1,2,3 \enspace \, , \\
&& \beta_{\bf 1} = \alpha + 2 K_{123}^{*}\lambda^{1}\lambda^{2}\lambda^{3} + \dots \, .
\end{eqnarray*}
At the third order the equation we need to check is
\begin{equation} \label{a}
\sum_{j=1}^{3}G_{ij}^{(1)}\beta^{j(1)} = r\frac{\partial \beta_{\bf 1}^{(3)}}{\partial \lambda^{i}} \, .
\end{equation}
(The term $(Z^{(3)})_{ren} - \beta^{i(1)}\partial_{i}(Z^{(3)})_{ren}$ vanishes similarly to the
$\Delta=1/2$ resonance  case above.)
 Using (\ref{G1}), (\ref{F}) we find that (\ref{a}) holds.

\section{Marginal but not exactly marginal couplings}
As a final case of universal beta function nonlinearities consider the case when marginal but not exactly
marginal couplings are present. Let us restrict our attention to a model situation when we have
perturbation by only three operators whose anomalous dimensions vanish:
$\epsilon_{1}=\epsilon_{2}=\epsilon_{3}=0$. Assume also for simplicity that only $C_{123}$ and its cyclic
permutations are nonvanishing.
 In that case we have
\begin{equation} \label{betas2}
\beta^{i} = \sum_{k,j=1}^{3}\frac{1}{2\pi}C_{ijk}\lambda^{j}\lambda^{k} + {\cal O}(\lambda^{3})
\end{equation}
with the first   term being scheme independent. It follows from (\ref{Z2}) that the $Z^{(2)}$ correction
to the partition function vanishes. The renormalized value of $Z^{(3)}$ can be computed the following way.
It is represented by the integral
\begin{eqnarray} \label{3tach}
&&Z^{(3)} = \lambda^{1}\lambda^{2}\lambda^{3}C_{(123)}K_{123} \, , \nonumber \\
&& K_{123} = \frac{1}{64\pi^{3}}\int\limits_{0}^{2\pi} \int\limits_{0}^{2\pi} \int\limits_{0}^{2\pi}
\frac{d\theta_{1}d\theta_{2}d\theta_{3}}{ |\sin\left( \frac{\theta_{1}-\theta_{2}}{2}\right)\sin\left(
\frac{\theta_{2}-\theta_{3}}{2}\right) \sin\left(\frac{\theta_{3}-\theta_{1}}{2}\right)|} \, .
\end{eqnarray}
The integral at hand is up to a constant factor  the integrated three-tachyon amplitude on a disk. Another
way of looking at such an integrated amplitude is that it gives the volume of  M$\ddot {\rm o}$bius group
${\rm PSL}(2, {\mathbb R})$. More precisely $K_{123} = \frac{1}{\pi^{3}}{\rm Vol}({\rm PSL}(2, {\mathbb
R}))$. Regulating the integral in (\ref{3tach}) by requiring that the distance between any two points is
greater than $\delta$ one finds that the M$\ddot {\rm o}$bius volume is linearly divergent
%% \footnote{The integral (\ref{3tach}) contains apparent
%logarithmic subdivergences when a pair of points comes together. The
%overall divergence that occurs when all 3 points come together has a piece that cancels logarithmic
%subdivergences and only a linearly divergent piece is left out.}
$$
{\rm Vol}({\rm PSL}(2, {\mathbb R})) = \frac{3\pi \ln 2}{\delta} - \frac{\pi^{2}}{2}
$$
and the renormalized value of the volume is $-\pi^{2}/2$ \cite{Tseytlin}, \cite{Liu_Polchinski}. Using
this value we obtain
$$
(Z^{(3)})_{ren} =-\frac{1}{2\pi}\lambda^{1}\lambda^{2}\lambda^{3}C_{(123)} \, .
$$
Using this expression, $G_{ij}^{(0)}=\delta_{ij}/2$ and formula (\ref{betas2}) we find that the gradient
formula considered at the first nonvanishing order
$$
G_{ij}^{(0)}\beta^{j(2)} = -\frac{\partial}{\partial \lambda^{i}}Z^{(3)}
$$
is correct.

\section{Some open questions}
Although  the gradient formula
(\ref{grad_f}), (\ref{g}),
(\ref{metric}) survived all the checks that  we performed above, that is certainly encouraging,
 many issues remain open. Below we point out to some of them.
 First note that in all our checks we employed a version of minimal
subtraction scheme that in general is known to behave very generously towards various
Ward identities. The precise scheme dependence of the gradient formula at hand still
needs to be clarified.

Another issue is the relation of this formula to the Affleck and Ludwig's computation \cite{AL2}. Consider
a cylinder partition function in the presence of non-scale-invariant boundary conditions. In this case we
can still define the boundary conditions via a boundary state $|B\rangle$ and the asymptotic
(\ref{cylinder_pf}) yields a quantity $\ln \langle B|0\rangle $ that in general contains a term
corresponding to a free-energy per unit length of the boundary
\begin{equation}\label{g(r)}
\ln \langle B|0\rangle = -rf_{B} + \log g(r) \, .
\end{equation}
The extensive free energy piece is non-universal and  for large $r$ dominates over the second piece. The
last one is believed to contain universal information and to interpolate between the UV $\ln g(0)$ and the
IR $\ln g(\infty)$ values of boundary entropy at the corresponding fixed points. While universality of
function $g(r)$ still remains a subtle issue (see \cite{DRTW} for some discussion) the free energy piece
certainly needs to be subtracted. The authors of \cite{AL2} carefully drop similar extensive terms in
their computation.

In the case of the gradient formula at hand as was already noted at the end of
section 2 the potential function can be represented in the form
\begin{equation} \label{gg}
g = Z - r\frac{d Z}{d r} \, .
\end{equation}
One may be tempted then to think that the role of the second term in (\ref{gg}) is
to subtract the boundary free energy extensive piece. For that to be the case
however one should have used $\ln Z$ in place of $Z$ in (\ref{gg})\footnote{
I would like to thank Alexander Zamolodchikov for stressing this point to me}.
 At the level of 3rd order computations  considered in this paper the only
difference between $Z$ and $\ln Z$ in (\ref{gg}) comes in the treatment of the identity coupling. It is
not hard to track that  to the given order in perturbation
nothing would change in the above checks of the
main gradient formula (\ref{grad2}) had we replaced $Z$ with $\ln Z$. However   equation
(\ref{grad_alpha}) for the derivative with respect to the identity coupling that previously was true
identically would stop holding. The whole issue  needs further clarification. It is not excluded that like
in the case of bulk gradient formula when various potential functions are possible \cite{CFL} (see also
\cite{Friedan} section 6.2) the boundary potential function (\ref{g}) and the function $g(r)$ in
(\ref{g(r)}) are essentially different off-criticality extensions of the boundary entropy.

Finally and  obviously a nonperturbative  proof of "g-theorem" remains to be  much desired.

\begin{center} {\bf\Large Acknowledgments} \end{center}

I am particularly grateful to Daniel Friedan for numerous discussions on the boundary gradient formula. It
is also a pleasure to thank Alexander Zamolodchikov for useful discussions and for reading the draft
version of the paper.

\appendix
\renewcommand{\theequation}{\Alph{section}.\arabic{equation}}
\setcounter{equation}{0}
\section{Two-point function}
In this appendix we give an explicit expression for the first order correction
 to the local two point function
in the minimal subtraction scheme.
It is given by the integral
\begin{eqnarray}
G_{12}(\theta_{12})=\sum_{k}\lambda_{k}r^{\epsilon_{1}+\epsilon_{2}+\epsilon_{k}}
\langle \phi_{1}(e^{i\theta_{1}})\phi_{2}(e^{i\theta_{2}})\int_{0}^{2\pi}
\frac{d\theta_{3}}{2\pi}\phi_{k}(e^{i\theta_{3}})\rangle = \nonumber \\
\sum_{k}\Bigl[4\sin^{2}\left(\frac{\theta_{12}}{2}\right)\Bigr]^{
\frac{1}{2}(\Delta_{k}-\Delta_{1}-
\Delta_{2})}\lambda_{k}r^{\epsilon_{1}+\epsilon_{2}+\epsilon_{k}}C_{12k}\times
\nonumber \\
 \frac{1}{2\pi}\int_{0}^{2\pi}d\phi
\Bigl[4\sin^{2}\left(\frac{\phi}{2}\right)\Bigr]^{\frac{1}{2}(\Delta_{1} -
\Delta_{2} - \Delta_{k})}
\Bigl[4\sin^{2}\left(\frac{\phi +\theta_{12}}{2}\right)\Bigr]^{\frac{1}{2}(
\Delta_{2} - \Delta_{1} - \Delta_{k})} \, .
\end{eqnarray}
The integral can be expressed via associated Legendre functions of the first kind
which in their turn are expressed via hypergeometric functions and after simplifications
we obtain
\begin{eqnarray} \label{G12}
&G_{12}(\theta_{12})=\sum_{k}\lambda_{k}(2r)^{\epsilon_{1}+\epsilon_{2}+\epsilon_{k}}
\frac{C_{12k}}{8\pi}
B(\frac{\epsilon_{2}+\epsilon_{k}-\epsilon_{1}}{2},
\frac{\epsilon_{1}+\epsilon_{k}-\epsilon_{2}}{2})
\nonumber \\
&\Bigl[\sin^{2}\left(\frac{\theta_{12}}{2}\right)\Bigr]^{(\epsilon_{1}+\epsilon_{2}+\epsilon_{k}
-2)/2}{}_{2}F_{1}\left(\frac{\epsilon_{2}+\epsilon_{k}-\epsilon_{1}}{2},
\frac{\epsilon_{1}+\epsilon_{k}-\epsilon_{2}}{2};\frac{1}{2};
1-\sin^{2}\left(\frac{\theta_{12}}{2}\right)\right)
\end{eqnarray}
where $B$ stands for the Euler's beta function.
In the resonance limit $\epsilon_{2}\to \epsilon_{1} + \epsilon_{k}$
we see the properly normalized logarithmic divergence. (This provides a simple check
of the answer.)

It is interesting to observe a jump in the short-distance behavior of (\ref{G12})
that happens at the value $\epsilon_{3}=1/2$: for $\epsilon_{3}<1/2$,
$G_{12}\sim (\theta_{12})^{\epsilon_{1}+\epsilon_{2} +\epsilon_{3}-1}$ while for
$\epsilon_{3}>1/2 $, $G_{12}\sim (\theta_{12})^{\epsilon_{1}+\epsilon_{2} -\epsilon_{3}}$.

\end{document}